
\input vanilla.sty

\def\frac#1#2{{#1\over#2}}
\def\<{\langle}\def\>{\rangle}
\def\Loop#1\Repeat{\global\n=0\global\let\body=#1\iterate}
\def\iterate{\body\let\next=\iterate\else\let\next=\relax\fi\next}
\def\ldd{\ifnum\n<\parenthesis\global\advance\n by1
\left.\nulldelimiterspace=0pt\mathsurround=0pt}
\def\rdd{\ifnum\n<\parenthesis\global\advance\n by1
\right.\nulldelimiterspace=0pt\mathsurround=0pt}
\def\nl{\Loop\rdd\Repeat\hfill\cr\qdd\Loop\ldd\Repeat{}}

\def\off#1{\hskip#1sp\relax}
\def\Nl{\hfill\cr}
\def\qdd{\quad\quad}
\def\nll{\hfil\newline}
\def\om{\Omega}
\def\ghp{{\cal P}}
\font\tenbf=cmbx10
\font\tenrm=cmr10
\font\tenit=cmti10
\font\ninebf=cmbx9
\font\ninerm=cmr9
\font\nineit=cmti9

\font\eightrm=cmr8
\font\eightit=cmti8
\font\sevenrm=cmr7
\TagsOnRight
\hsize=5.0truein
\vsize=7.7truein
\parindent=15pt
\nopagenumbers
\vglue 5pc
\baselineskip=13pt
\centerline{\tenbf COMPUTING THE BRST OPERATOR USED }
\baselineskip=15pt
\centerline{\tenbf IN QUANTIZATION OF GAUGE THEORIES}
\vglue 1cm
\centerline{\eightrm A. BURNEL, H. CAPRASSE%
\footnote "$^{\star}$"{\eightrm\baselineskip=10pt E-mail:
caprasse\@vm1.ulg.ac.be}}
\baselineskip=12pt
\centerline{\eightit D\'epartement d'Astronomie et d'Astrophysique,
Universit\'e de Li\`ege}
\baselineskip=10pt
\centerline{\eightit Institut de Physique (B5),
B-4000 Li\`ege I, Belgium}
\vglue 0.2cm
\centerline{and}
\vglue 0.2cm
\centerline{\eightrm A. DRESSE%
\footnote "${^+}$" {\eightrm\baselineskip=10pt E-mail: adresse\@ulb.ac.be}}
\baselineskip=12pt
\centerline{\eightit Universit\'e libre de Bruxelles,
  CP 210/01 Blvd du Triomphe}
\baselineskip=10pt
\centerline{\eightit  B-1050 Bruxelles }
\vglue 1cm
\centerline{\eightrm ABSTRACT}
{\rightskip=1.5pc
 \leftskip=1.5pc
 \eightrm\baselineskip=10pt\parindent=1pc
It is shown that for a large class of non-holonomic quantum mechanical
systems one can make the computation of BRST charge fully
algorithmic. Two computer algebra programs written in the language of
{\tt REDUCE} are described. They are able to
realize the complex calculations  needed
to determine the charge for general nonlinear algebras.
Some interesting specific  solutions are discussed.
\vglue 5pt
\noindent
{\eightit Keywords}\/:{ Gauge Theory; Computer Algebra.}
\vglue 12pt}
\vglue 12pt
\baselineskip=13pt
\line{\tenbf 1. Introduction \hfil}
\vglue 0.2cm
BRST theory$^1$ has proven to be a very powerful tool to treat
gauge theories and, in particular, to quantize gauge  field theories.
Indeed it may be useful to get the physical states from a space larger than
seems necessary, because the restriction of the phase space to the
physically meaningful configurations often hides the natural
symmetries. Some systems, such as Yang-Mills theory, do not even admit
global gauge fixing conditions.
In such cases BRST
theory naturally introduces the ghosts, necessary to write the path integral
formulation of gauge theories.

At the core of BRST theory is the BRST charge $Q_B$, used to select the
physical states from a larger space, and specify the physical equivalence
of two a priori different states.
$$
\align
& Q_{B} |\Psi_{phys}> =0, \, \, \tag 1.1 \\
& |\Psi_{phys}>\quad\approx\quad |\Psi'_{phys}> +Q_{B}|\Psi_0>.\, \, \tag 1.2
\endalign
$$

The BRST charge is closely related to the symmetries of the system
described in the Lagrangian formulation through Noether theorem.
Its construction can, however, be based only on the knowledge of the algebra
of constraints.

A recursive procedure to build the BRST operator order by order in the ghosts
has been developped but that procedure is not fully algorithmic. We show here
that the procedure can be made algorithmic for an important class of theories.
We describe the capabilities of two programs written in {\tt REDUCE$^2$}
which do indeed allow
to compute
the corresponding BRST charge.
This class covers quadratic algebras such as those discussed in ref.~3,4, thus
extending the realm of application of BRST theory beyond usual gauge theories.
The use of computer algebra is mandatory for all but the most simple cases
as the computations involved are tedious, although systematic.

The paper is organized as follows.\nll In sec.~2, the construction of the
BRST operator is briefly described and two algorithms for its construction are
given.
The programs are described in sec.~3 while their applications
are illustrated in sec.~4. Conclusions as well as the present limitations
of these programs are presented in sec.~5.
\vglue 0.5cm
\line{\tenbf 2. The BRST operator \hfil}
\vglue 0.2cm
In the Hamiltonian formulation of a gauge theory, the symmetries of the
system are described by the constraints, which are relations between the
generalized coordinates and the momenta. The constraints of the Hamiltonian
formulation are not all related to gauge invariance. Those related to it
are
called first class constraints. The others, said to be of second class,
can be eliminated by an adequate redefinition of the brackets.

Let us consider a set of $N$ first class constraints
$$
\{C_a ,\quad a =1,\ldots , N\}.
$$
They satisfy a Poisson bracket algebra
$$
\{C_a ,C_b \}=f_{ab}^{\ \ c} C_c \, \, \tag 2.1
$$
where the f's, the structure functions depend on the canonical variables.
Here and in the following, summation over repeated indices is understood
except otherwise stated.

The computation of the BRST charge can be made fully algorithmic if we assume
that one can write the structure functions in terms of the constraints.
Eq (2.1) can then be written as
$$
\{C_a ,C_b \}=\sum_{i=1}^q f_{i\, ab}^{a_1\ldots a_i}
\prod_{j=1}^i C_{a_j} \, \,  \tag 2.2
$$
where, now, the $f_i$'s are constants and $q$ is a number
obtained from  the relations between the constraints and the structure
functions. In the present work, the necessary manipulations to obtain it
are not considered so that Eq.~(2.2) is the starting point.

To each constraint $C_a$,
one associates two variables of odd Grassmann parity :
a ghost $\eta^a$ with ghost number 1
and a ghost momentum ${\cal P}_a$ with ghost number -1.
They satisfy the following bracket algebra\footnote"$^a$"
{\eightrm\baselineskip=10pt The choice of sign in the bracket
of ghosts and ghost momenta is conventional. In this paper,
the conventions of ref.~1 are adopted.}.

$$
\{\eta^a ,\eta^b\}=\{{\cal P}_a ,{\cal P}_b\}=0,\quad
\{\eta^a ,{\cal P}_b\}=-\delta_b^a. \, \, \tag 2.3
$$

The BRST operator $\Omega$ is defined by
\halign{ \qquad \qquad \qquad # \hfil \cr
- its nilpotency ($\{\Omega,\Omega\}=0$), \cr
- its ghost number 1 \cr
- it involves the term $\eta^a C_a$ \cr
}
\vglue .1cm

It can be decomposed in powers of the ghosts as
$$
\Omega =\sum_{n=0}^{N-1}\Omega^{(n)}=
\sum_{n=0}^{N-1} A_{i_1\ldots i_{n+1}}^{j_1\ldots j_n}
\eta^{i_1}\ldots\eta^{i_{n+1}}
{\cal P}_{j_1}\ldots{\cal P}_{j_n} \, \, \tag 2.4
$$
where the $A$'s depend only on the constraints and structure constants
of the problem. It can be determined in a recursive way by
$$
\align
\Omega^{(0)}&=\eta^a C_a, \, \, \tag 2.5 \\
\delta\Omega^{(n+1)}&=-D^{(n)} \, \, \tag 2.6
\endalign
$$
with
$$
\delta F=-F{\overleftarrow\partial\over\partial{\cal P}_a}\,
      C_a \, \, \tag 2.7
$$
and
$$
D^{(n)}={1\over2}\left( \sum_{k=0}^n\{\Omega^{(k)},\Omega^{(n-k)}\}_C
+ \sum_{k=0}^{n-1}\{\Omega^{(k+1)},\Omega^{(n-k)}\}_{gh}\right). \tag 2.8
$$
The second term is not present when $n=0$.
Here, $\{\cdot,\cdot\}_C$ and $\{\cdot,\cdot\}_{gh}$ are respectively
the constraint and ghost Poisson brackets.

The computable  $D$ is written in the form
$$
D^{(n-1)}=Z^{(a)a_1\cdots a_{n-2}b}C_{(a)}{\cal P}_{a_1}\cdots
{\cal P}_{a_{n-2}}C_b
\,\,\tag 2.9
$$
in order to make the computation algorithmic.
The $Z$'s depend on the ghosts $\eta^a$ and structure constants and
are antisymmetric over the $a_1,\cdots, a_{n-2},b$.
$(a)$ is a set of indices which may be empty.
Note that this expression is not easy to obtain explicitly, as it involves
a decomposition of a polynomial ($D^{(n-1)}$) in  products of the constraints,
seen as generators of a polynomial ideal.

{}From Eq. (2.4), one writes
$$
\Omega^{(n)}=K^{(a)a_1\cdots a_{n-1}}C_{(a)}{\cal P}_{a_1}
\cdots {\cal P}_{a_{n-1}}
\,\,\tag 2.10
$$
where $K^{(a)a_1\cdots a_{n-1}}$ is
antisymmetric over $a_1,\cdots ,a_{n-1}$.
$C_{(a)}$ denotes the product of constraints whose indices are
in the set $(a)$.

Eq.~(2.6) leads to
$$
(n-1)K^{(a)a_1\cdots a_{n-2}b}C_{(a)}{\cal P}_{a_1}
\cdots {\cal P}_{a_{n-2}}C_b
=Z^{(a)a_1\cdots a_{n-2}b}C_{(a)}{\cal P}_{a_1}
\cdots {\cal P}_{a_{n-2}}C_b.
\,\,\tag 2.11
$$
Its solution is
$$
K^{(a)a_1\cdots a_{n-2}b}={1\over n-1}
\left( Z^{(a)a_1\cdots a_{n-2}b} +G^{(a)a_1\cdots a_{n-2}b}\right)
\,\,\tag 2.12
$$
as Z is antisymmetric as mentionned.

The tensor $G$ is arbitrary except that it is
antisymmetric over the set of indices formed by the union of
$a_1,\cdots ,a_{n-2},b$ and one index of the set $(a)$.
In practice, it will be built from the tensor $Z$  with the above
antisymmetrisation, each possible term being multiplied by an arbitrary
coefficient.

Since the BRST charge is defined only up to an arbitrary BRST exact
term, there exist infinitely many solutions of Eq.~(2.6).
In particular, the $\Omega^{(n)}$ can be built as an arbitrary
linear combination of the type
$$
\sum_{i=0}^{n-1}\alpha_i Z^{a_1\ldots a_{i-1} (a) a_i\ldots a_{n-1}}
\,\,\tag 2.13
$$
where no antisymmetrisation on the indices of $Z$ is involved.
Some of the coefficients $\alpha_i$ will be fixed
by imposing Eq.(2.6). It is left to find heuristic criteria to fix the
remaining ones. Two of them are explained in the next section together
with the corresponding programs.
\vglue .5cm
\line{\tenbf 3. Description of the Programs \hfil}
\vglue 0.2cm
A preliminary step to the writing of the programs is to find ways to
fix all constants $\alpha_i$'s. This is done in two ways. They are successively
described.

In the first one, during the calculation of each $\Omega^{(n)}$, the
coefficients $\alpha_i$'s are chosen to minimize the number of terms.

In the second one, one assumes that the BRST charge remains invariant
under a redefinition of the constraints, of the ghosts and the ghost momenta.
This kind of constraint is usual in homological perturbation theory.
The principles at the basis of this choice are the following:

Suppose one can define a linear operator $\sigma$, and a partition
${\cal A} = \oplus_k {\cal A}_k$ of the algebra of the constraints
such that, for all $n$,
$$
(\delta \sigma + \sigma \delta) A = N_n A\, \, \tag 3.1
$$
for all $A \in {\cal A}_n$. Such an operator is called a contracting
homotopy operator for $N$.

Now consider the equation (2.6)
where $D^{(n)}$ is known, and $\Omega^{(n+1)}$ has to be determined.
Decompose them as
$$
\align
  \Omega^{(n+1)} &= \sum_i \Omega^{(n+1)}_i \qquad \Omega^{(n+1)}_i
  \in {\cal A}_i,\, \, \tag 3.3  \\
  D^{(n)} &= \sum_i  D^{(n)}_i \qquad D^{(n)}_i \in {\cal A}_i.\, \, \tag 3.4
\endalign
$$

$\delta D^{(n)}_i = 0$ since $\delta D^{(n)} = 0$ and $\delta$ is linear.
Thus, solving Eq. (2.6) is equivalent to solving
$\delta \Omega^{(n+1)}_i = -D^{(n)}_i$ for all $i$, as
$$
  \delta \Omega^{(n+1)}_i = -D^{(n)}_i \Rightarrow
  \delta \Omega^{(n+1)} = \delta \sum_i \Omega^{(n+1)}_i =
  \sum_i \delta \Omega^{(n+1)}_i = - \sum_i D^{(n)}_i = -D^{(n)}.\, \, \tag 3.5
$$
Now, since
$$
  \delta \sigma D^{(n)}_i = (\delta \sigma + \sigma \delta)
  D^{(n)}_i = N_i D^{(n)}_i, \, \, \tag 3.6
$$
as $\delta D^{(n)}_i = 0$, one sees that
$$
  \Omega^{(n+1)}_i = {1 \over {N_i}} \sigma D^{(n)}_i\, \, \tag 3.7
$$
is effectively a solution at degree $i$.

One can then state that for a contracting homotopy $\sigma$ for $N$,
$$
  \Omega^{(n+1)} = \sum_i {\frac{1}{N_i}} \sigma D^{(n)}_i \, \, \tag 3.8
$$
is a solution of the equation (2.6).

A particular contracting operator $\sigma$ is given by
$$
\sigma = {{\overleftarrow \partial} \over {\partial C_a}} {\cal P}_a
\, \, \tag 3.9
$$
where the derivative with respect to the constraints is defined
since the expressions considered depend only on the phase space
variables through the constraints.

This definition of the contracting operator corresponds to a particular
choice of the values of the parameters $\alpha_i$. Other choices would
result in other partitions of the set of expressions considered, and
other contracting operators.

The BRST charge resulting from the current choice
of contracting homotopy operator is invariant under linear redefinitions
of the constraints:
$$
C_a \rightarrow C'_a = A_a{}^b C_b \, \, \tag 3.10
$$
with the appropriate redefinitions of the ghosts and ghost momenta:
$$
\align
{\cal P}_a &\rightarrow {\cal P'}_a = A_a{}^b {\cal P}_b \\
\eta^a &\rightarrow {\eta'}^a = (A^{-1})_b{}^a \eta^b. \, \, \tag 3.11
\endalign
$$

One is ready now to describe the two programs computing the BRST charge.
They are based on the two choices of constants $\alpha_i$ 's described above.
On the other hand, one is written in the algebraic mode of {\tt REDUCE}, the
other is written in its symbolic mode. A subsidiary usefulness of writing
two programs is
to control the validity of the calculations.
Apart from the algorithms, their originality does not come from the handling
of anticommuting variables (the ghost and ghost momenta) but from the extended
use of dummy indices to represent the various expressions.
Thanks to that, one is able to
apply them without reference to the number of constraints and, of course,
the complexity of the computations is also independent on this number.
Expressing the brackets $\{\eta^a C_a, X\}$ for example
involves as many calculations as there are constraints if the summation
is explicit. On the other hand, non explicit summations on dummy variables
allow the treatment of generic cases. However, the problem is now to
achieve full simplification of polynomials.
Outside the context of tensor algebra, there
exist no package which can do that except the package
{\tt DUMMY$^5$} recently created by one of the authors.
Both programs also use the package {\tt ASSIST$^6$} but they use
different functionalities included in it.
\vglue .3cm
{\tenbf 3.1. Program I}
\vglue .2cm
{\ninebf Input}:
\halign{ \qquad \qquad \qquad # \hfil \cr
- Algebra of constraints Eq. (2.2) \cr
- $\Omega^{(0)}$ Eq. (2.5) \cr}
{\ninebf Output}: $\Omega^{(n)}$
\vglue .1cm
The process of computations is:
\halign{ \qquad \qquad \qquad # \hfil \cr
\quad  $k=0$ \cr
LOOP: \cr
\qquad compute $D^{(k)}$ from Eq. (2.8) \cr
\qquad extract the corresponding $Z$  from Eq. (2.9) \cr
\qquad construct $\Omega^{(k+1)}$ from Eqs. (2.12) and (2.10) \cr
\qquad arbitrary coefficients left in solution
optionally fixed as explained below \cr
\qquad if $k=n-1$ then return $\Omega^{(n)}$  else k:=k+1 go to LOOP \cr}
\vglue .1cm
In the program one has to handle polynomials of the type
$$
\sum a_{b_1\ldots b_m}^{a_1\ldots a_lc_1\ldots c_n}
C_{a_1}\ldots C_{a_l} \eta^{b_1}\ldots\eta^{b_m}
{\cal P}_{c_1}\ldots{\cal P}_{c_n}
\,\,\tag 3.12
$$
where $\eta$ and ${\cal P}$ are odd Grassmann variables and the $a$'s
stand for $Z$ or $K$. It is useful
to consider the ghost and ghost momenta products as antisymmetric
operators :
$$\eta^{b_1}\ldots\eta^{b_m} =\eta^{b_1\ldots b_m},\quad
{\cal P}_{c_1}\ldots{\cal P}_{c_n}={\cal P}_{c_1\ldots c_n}. \, \, \tag 3.13
$$
\vglue .3cm
\line{\ninebf a. calculation of $D$\hfill}
\vglue .2cm
Brackets of expressions like (3.12)
can be decomposed into two parts, the constraint brackets
$$
\sum\sum
a_{b_1\ldots b_m}^{a_1\ldots a_l c_1 \ldots c_n}
b_{b'_1\ldots b'_{m'}}^{a'_1\ldots a'_{l'} c'_1\ldots c'_{n'}}
\{C_{a_1}\ldots C_{a_l},C_{a'_1}\ldots C'_{a_{l'}}\}_C
$$
$$
\eta^{b_1\ldots b_m b'_1\ldots b'_{m'}}
{\cal P}_{c_1\ldots c_n c'_1\ldots c'_{n'}} (-1)^{nm'}\, \, \tag 3.14
$$
and the ghost brackets
$$
\sum\sum
a_{b_1\ldots b_m}^{a_1\ldots a_lc_1\ldots c_n}
b_{b'_1\ldots b'_{m'}}^{a'_1\ldots a'_{l'} c'_1\ldots c'_{n'}}
C_{a_1}\ldots C_{a_l}C_{a'_1}\ldots C_{a_{l'}}
$$
$$
\{\eta^{b_1\ldots b_m}{\cal P}_{c_1\ldots c_n},
\eta^{b'_1\ldots b'_{m'}}{\cal P}_{c'_1\ldots c'_{n'}}\}.\, \, \tag 3.15
$$
In both cases, the time of calculation is considerably shortened if either
the $C$'s or the ghosts or both factorize.
In practice, a factorization of the ghosts
is more frequent than a factorization of the constraints.

Using the Leibniz rule, the constraint brackets are easily computed from
the input.

The ghost brackets implement the formula

$$
\left\{\eta^{a_1\ldots a_{i_1}}{\cal P}_{b_1\ldots a_{i_2}},
\eta^{a '_1\ldots a '_{j_1}}
{\cal P}_{b '_1\ldots b '_{j_2}}\right\} =
$$
$$\left[
 \sum_{k=1}^{i_1}\sum_{l=1}^{j_2}
\eta^{a_1\ldots a_{k-1}a_{k+1}\ldots a_{i_1}a '_1\ldots
a '_{j_1}}
{\cal P}_{b_1\ldots b_{i_2}b '_1\ldots b '_{l-1}b '_{l+1}
\ldots b '_{j_2}}
(-1)^{k+l+1+i_2j_1}\delta_{b '_l}^{a_k}\right. +
$$
$$
\left.
\sum_{k=1}^{j_1}\sum_{l=1}^{i_2}
\eta^{a_1\ldots a_{i_1}a '_1\ldots a'_{k-1}
a'_{k+1}\ldots a '_{j_1}}
{\cal P}_{b_1\ldots b_{l-1}b_{l+1}\ldots b_{i_2}
b '_1\ldots b '_{j_2}}
\delta_{b_l}^{a '_k}(-1)^{k+l+j_1j_2+i_1i_2+i_2j_1}\right].
$$
$$\, \, \tag 3.16
$$
This formula can be checked by any program dealing
with anticommutating variables.
The $D$ function is obtained from the bracket calculation.
Because it involves many terms with the same structure,
the use of the function {\tt CANONICAL} from the package
{\tt DUMMY}
is essential to simplify it.
Because $\delta D^{(n)}=0$, one checks this vanishing for each value of $n$.
It is interesting to note that,
in all the considered examples,
$\delta D^{(1)}=0$ gives all the Jacobi
identities for the structure constants of the algebra.
\vglue .3cm
\line{\ninebf b. determination of the $Z$ function\hfill}
\vglue .2cm
Because the result for D is obtained from
{\tt CANONICAL}, an antisymmetrization over the ghost indices is necessary
to extract $Z$. After
this antisymmetrization, each monomoial is divided by the constraints and the
ghost momentum operator. This gives  $Z$.
\vglue .3cm
\line{\ninebf c. construction of the BRST operator \hfill}
\vglue .2cm
It is constructed from Eq. (2.12) where  the arbitrary $G$'s are built from
$Z$ by a further antisymmetrization and multiplication by
arbitrary coefficients. These appear in the output but can also be fixed
if one requires the number of terms appearing in the output to be minimal.
\vglue .3cm
\vfil\newpage
{\tenbf 3.2. Program II}

The input and output for the second program are identical to that of
the first program:
{\ninebf Input}:
\halign{ \qquad \qquad \qquad # \hfil \cr
- Algebra of constraints Eq. (2.2) \cr
- $\Omega^{(0)}$ Eq. (2.5) \cr}
{\ninebf Output}: $\Omega^{(n)}$
\vglue .1cm
The process of computation is also similar, as it follows the
standard steps described earlier.

The BRST charge computed here differs from that returned by
the first program in the choice of values for the coefficients $\alpha$.
We have indeed chosen to adopt here the contracting operator (3.9)
which yields a BRST charge invariant under linear transformations
of the constraints.

This choice induces a larger number of terms in the expression of
$\Omega$, and thus requires more computer resources. It was
therefore necessary to work in the symbolic mode of REDUCE.

A further particularity of this program compared to the previous one
lies in its handling of the Jacobi identity. Indeed, these were useful
in reducing as much as possible the number of terms in the expressions
without changing the contracting homotopy operator. We have built
a procedure returning a normal form of polynomials with respect to the
Jacobi identity for the particular algebras studied. This enabled us
not only to reduce the size of the expressions, but also to check the
validity of the results returned.

It should be noted however that this handling of side relations is
very time consuming, and requires {\it ad hoc} tweaking for each new algebra.

Finally, this program has been used to study a partial classification of
polynomial Poisson structures$^7$.
\vglue 0.5cm
\line{\tenbf 4. Results \hfil}
\vglue 0.2cm
Various algebras have been considered
\vglue .3cm
\line{\ninebf Usual linear Lie algebras\hfill}
\vglue .2cm
The constraint algebra is given by
$$\{C_a,C_b\}=f_{ab}^{\ \ c}C_c.\, \, \tag 4.1
$$
Use of both programs gives
$$\Omega^{(1)}={1\over2}f_{ab}^{\ \ c}\eta^{ab}{\cal P}_c.\, \, \tag 4.2
$$
Computing $D^{(1)}$ and checking the $\delta D^{(1)}=0$
lead to the Jacobi identity
$$f_{[ab}^c\, f_{d]c}^e =0.\, \, \tag 4.3
$$
This leads to
 $D^{(1)}=0$
and stops the construction because all the added contributions vanish.

\vglue .3cm
\line{\ninebf Self-reproducing algebras\hfill}
\vglue .2cm
Let us consider the algebra
$$\{C_a,C_b\}=T_{ab}C_aC_b\, \, \tag 4.4
$$
where no summation over a,b is involved. In this section there is no
summation over repeated indices.
Such an algebra is characterized by the fact that Jacobi identities are
trivial.
At each order from the second one, an arbitrary coefficient is generated.
The number of handled terms increase quickly. Fortunately, a particular
choice of the arbitrary coefficient $\alpha_n$ occuring at the order $n$ by
$$\alpha_n=-{n+1\over n}\, \, \tag 4.5
$$
considerably simplifies the result and the BRST operator
can be obtained with program I\ in a closed form.
$$\Omega=\sum_{n=0}^{N-1} {(-1)^{n(n-1)\over 2}\over 2^n n!}
\sum_b (K_b)^n C_b\eta^b\, \, \tag 4.6
$$
where
$$K_b=\sum_a\eta^a {\cal P}_a T_{ab}\, \, \tag 4.7
$$
The program is also able to compute $\Omega$ for arbitrary coefficients.
An example of output is given in Appendix A. The expressions are much more
complicated but they allow to check the compatibility of the calculations
made by both programs.
\vglue .3cm
\line{\ninebf Pure quadratic algebras\hfill}
\vglue .2cm

The BRST operator has been computed for the constraint algebra
$$\{C_a,C_b\}=d_{ab}^{cd}C_cC_d\, \, \tag 4.8
$$
along the same lines as above. One gets
$$\Omega^{(1)}={1\over2}d_{ab}^{cd}\eta^{ab}{\cal P}_cC_d.\, \, \tag 4.9
$$
The vanishing of  $\delta D^{(1)}$ gives the Jacobi identity
$$d_{[ab}^{c\{g}\,  d_{d]c}^{ef\}} =0.\, \, \tag 4.10
$$
The calculation of $\Omega^{(2)}$ gives
$$\Omega^{(2)}={1\over6}d_{ab}^{cd}d_{ec}^{fg}\eta^{abc}{\cal P}_{df}C_g.
\, \, \tag 4.11
$$
Program II\ allows to easily compute $\Omega$ up to order six.
\vglue .3cm
\line{\ninebf Mixed linear and quadratic algebras\hfill}
\vglue .2cm
The following algebra
$$\{C_a,C_b\}=f_{ab}^{\ \ c}C_c + d_{ab}^{cd}C_cC_d\, \, \tag 4.12
$$
with
$$d_{ab}^{cd}d_{ce}^{fg}=0\tag 4.13 $$
is an extension of an algebra studied by Schoutens, Sevrin and
Van Niewenhuizen$^4$.
Applying the programs, one gets
$$\Omega^{(1)}={1\over2}\left(
f_{ab}^{\ \ c}\eta^{ab}{\cal P}_c +
d_{ab}^{cd}\eta^{ab}{\cal P}_cC_d\right).\, \, \tag 4.14
$$
The vanishing of $\delta D^{(1)}$ again leads to the  Jacobi identities
$$f_{[ab}^{\ \ c}f_{d]c}^{e}=0,\, \, \tag 4.15
$$
$$f_{[ab}^{\ \ c}d_{d]c}^{ef}+d_{[ab}^{ce}f_{d]c}^{\ \ f}+
d_{[ab}^{cf}f_{d]c}^{\ \ e}=0.\, \, \tag 4.16
$$
Together with (4.13), they imply  $D^{(1)}=0$ and $\Omega^{(2)}=0$.

$\Omega^{(3)}$ is given by
$$\Omega^{(3)}={1\over24}d_{ab}^{pe}d_{cd}^{qf}f_{pq}^{\ \ g} \eta^{abcd}
{\cal P}_{efg}.\, \, \tag 4.17
$$

One can note that here and in the previous example, no arbirary coefficient
is involved.
The condition (4.13) implies the vanishing of higher orders. If this condition
is released one can again compute easily $\Omega$ up to order six with
program~II.
\vglue .3cm
\line{\ninebf An example of cubic algebra\hfill}
\vglue .2cm
The  cubic algebra generated by the $C$'s


$$
\{C_{d_1}, C_{d_2}\} = f_{d_1 d_2}^{d_3} C_{d_3} +
    D_{d_1 d_2}^{d_3 d_4} C_{d_3} C_{d_4} +
     E_{d_1 d_2}^{d_3 d_4 d_5} C_{d_3} C_{d_4} C_{d_5}
\, \, \tag 4.18
$$
is an extension of the wellknown spin 4 algebra.
The various structure constants $f,D,E$ are antisymmetric over
their lower indices and symmetric over their upper indices.
They satisfy the Jacobi identities~:
$$
\align
& f_{[ab}^{\ \ c}f_{d]c}^{e}=0, \, \, \tag 4.19\\
& f_{[ab}^{\ \ c}d_{d]c}^{ef}+D_{[ab}^{ce}f_{d]c}^{\ \ f}+
D_{[ab}^{cf}f_{d]c}^{\ \ e}=0, \, \, \tag 4.20\\
& 2D_{d_7[d_4}^{\{d_1d_2}D_{d_5d_6]}^{d_3\}d_7} +
E_{d_7[d_4}^{\{d_1d_2d_3\}}f_{d_5d_6]}^{d_7} +
3f_{d_7[d_4}^{\{d_1}E_{d_5d_6]}^{d_2d_3\}d_7}=0, \, \, \tag 4.21\\
& 3D_{d_1[d_2}^{\{d_3d_4}E_{d_7d_8]}^{d_5d_6\}d_1}+
2E_{d_1[d_2}^{\{d_3d_4d_5}D_{d_7d_8]}^{d_6\}d_1}=0, \, \, \tag 4.22\\
& E_{[d_6d_7}^{d_8\{d_1d_2}E_{d_9]d_8}^{d_3d_4d_5\}}=0. \, \, \tag 4.23
\endalign
$$
The BRST charge for this algebra is given to order six in Appendix B.
\vglue 0.5cm
\line{\tenbf 5. Conclusions. \hfil}
\vglue 0.2cm
As shown by the results in the previous sections, the new programs
allow the computation of the BRST operator when the algebras of constraints
are more complicated than the usual linear algebras. We recall that
quadratic algebras have been considered recently in the framework of
the study of superconformal field theories$^4$
Program I is written in the algebraic mode of {\tt REDUCE} . It is,
of course, less efficient than the program written in symbolic mode but is
still quite able to make the most relevant computations in a reasonable
time. Its main limitation is the fact that it does not take
properly into account the Jacobi identities. For instance, the
calculation of the first nontrivial term after $\Omega^{(1)}$ is incorrect
by a numerical factor ${1\over3}$. The necessity to take into account the
Jacobi identities  can be implemented in program I but
program II fulfills that job and, so it does not look to be worth the task.
As far as this aspect is concerned, it should be stressed that it is
because of the high level of
symmetry of Poisson algebra structures that this was possible and
it is not claimed that the algorithm is efficient.
Finally, the specificity of program I with respect to program II lies in the
calculation of the BRST operator in self-reproducing algebras.
The result can indeed be written in a more compact form than the choice of
the contracting homotopy made in program~II allows to.
Programs are available, on request, by  electronic mail at the following
address caprasse\@vm1.ulg.ac.be.
\vglue 0.5cm
\line{\tenbf References \hfil}
\vglue 5pt
\medskip
\ninerm
\baselineskip=11pt
\frenchspacing
\item{1.} M.~Henneaux and C.~Teitelboim, {\nineit Quantization of
Gauge Systems}, Princeton University Press, New Jersey (1993).
\item{2.} A.C.~Hearn, {\nineit ``User's Manual'', Version 3.5}
RAND, Santa Monica, CA 90407-2138J. (1993).
\item{3} K.~Schoutens, A.~Sevrin, P.~van Nieuwenhuizen,
{\nineit Commun. Math. Phys. {\ninebf 124}, 87 (1989)}
\item{4} Z. Khviengia and E. Sezgin, {\nineit Phys. Lett. {\ninebf B 326},
243 (1994)}
\item{5} A.~Dresse {\eightit ``DUMMY.RED''}\ \ {\tt REDUCE}\ library (1994).
\item{6} H.~Caprasse {\eightit ``ASSIST.RED''}\ \ {\tt REDUCE}\ library (1993).
\item{7} A.~Dresse and M.~Henneaux {\nineit J. Math. Phys. {\ninebf 35 vol. 3},
1334 (1994)}
\item{8} W.~Antweiller {\eightit ``TRI.RED''}\ \  {\tt REDUCE}\ library (1993).
\vfill\newpage
\vglue .5cm
\baselineskip=10pt
\line{\tenbf Appendix A \hfill}
\vglue .3cm
\noindent
$omega(3)$ computed from program I for arbitrary parameters $alpha2,
alpha3$ :
\vglue .3cm
\line{$omega(3):=brstconstr(3,any);$ \hfill}
$$
\align
omega(3) & := (alop(s1,s2)*alop(s1,s3)*eta(s1,s2,s3,s4)* \\
& (- 3*alop(s1,s4)*contr(s1)*prond(s2,s3,s4)*alpha2*alpha3 \\
&  - 3*alop(s1,s4)*contr(s1)*prond(s2,s3,s4)*alpha3 \\
&  + 9*alop(s1,s4)*contr(s4)*prond(s1,s2,s3)*alpha2*alpha3 \\
&  + 12*alop(s1,s4)*contr(s4)*prond(s1,s2,s3)*alpha2 \\
&  + 9*alop(s1,s4)*contr(s4)*prond(s1,s2,s3)*alpha3 \\
&  + 12*alop(s1,s4)*contr(s4)*prond(s1,s2,s3) \\
&  + 56*alop(s3,s4)*contr(s3)*prond(s1,s2,s4)*alpha2 \\
&  + 84*alop(s3,s4)*contr(s3)*prond(s1,s2,s4) \\
& + 8*alop(s3,s4)*contr(s4)*prond(s1,s2,s3)*alpha2 \\
& + 12*alop(s3,s4)*contr(s4)*prond(s1,s2,s3)))/96\$
\endalign
$$
$omega(3)$ computed from program I with $alpha2=-3/2,\; alpha3=-4/3$ :
\vglue .3cm
\line{$omega(3):=brstconstr(3,simplify);$ \hfill}
$$
\align
omega(3) & := (alop(s1,s2)*alop(s1,s3)*alop(s1,s4)*alop(s1,s5)*alop(s1,s6)* \\
& contr(s1)*eta(s1,s2,s3,s4,s5,s6)*prond(s2,s3,s4,s5,s6))/3840
\endalign
$$
\vfill
\newpage
\newcount\parenthesis \parenthesis=0 \newcount\n
\def\({\global\advance\parenthesis by1\left(}
\def\){\global\advance\parenthesis by-1\right)}
\def\{{\global\advance\parenthesis by1\left\lbrace}
\def\}{\global\advance\parenthesis by-1\right\rbrace}
\def\[{\relax} 
\def\]{\relax} 
\baselineskip=10pt
\vglue .3cm
\line{\tenbf Appendix B \hfill}
\vglue 0.5cm
\noindent
An example of the output from Program II for the cubic algebra follows.
The formula below is a direct \TeX\ output of package {\tt TRI}$^8$
of {\tt REDUCE}. It should be
clear that we can compute $\om$ to higher orders.
$$\displaylines{\qdd
\om(6)=
\[\(\eta^{d_{1}}\,\eta^{d_{2}}\,\eta^{d_{3}}\,
    \eta^{d_{4}}\,\eta^{d_{5}}\,\eta^{d_{6}}\,
    \eta^{d_{7}}\,\ghp_{d_{8}}\,\ghp_{d_{9}}\,
    \ghp_{d_{10}}\,\ghp_{d_{11}}\,\ghp_{d_{12}}\,
    \ghp_{d_{13}}\,\nl
    \off{3499956}
    \(864\,f_{d_{14}d_{2}}^{
              d_{9}}\,f_{d_{15}d_{16}}^{
                         d_{10}}\,f_{d_{17}d_{18}}^{
      d_{11}}\,C_{d_{19}}\,E_{d_{3}d_{4}}^{
                              d_{15}d_{17}d_{12}}\,
      E_{d_{5}d_{6}}^{
         d_{18}d_{19}d_{13}}\,\nl
      \off{3827636}
      E_{d_{7}d_{1}}^{
         d_{14}d_{16}d_{8}}
      +864\,f_{d_{14}d_{15}}^{
               d_{9}}\,f_{d_{16}d_{2}}^{
                          d_{10}}\,f_{d_{17}d_{18}}^{
      d_{11}}\,C_{d_{19}}\,E_{d_{3}d_{4}}^{
                              d_{17}d_{19}d_{12}}\nl
      \off{3827636}
      \,E_{d_{5}d_{6}}^{
           d_{14}d_{16}d_{13}}\,E_{d_{7}d_{1}}^{
      d_{15}d_{18}d_{8}}
      -126\,f_{d_{14}d_{15}}^{
               d_{9}}\,f_{d_{16}d_{17}}^{
                          d_{10}}\,D_{d_{3}d_{4}}^{
      d_{18}d_{12}}\,\nl
      \off{3827636}
      D_{d_{18}d_{2}}^{
         d_{16}d_{11}}\,C_{d_{19}}\,E_{d_{5}d_{6}}^{
      d_{14}d_{17}d_{13}}\,E_{d_{7}d_{1}}^{
                              d_{15}d_{19}d_{8}}
      -63\,f_{d_{14}d_{15}}^{
              d_{9}}\,D_{d_{2}d_{3}}^{
                         d_{16}d_{10}}\nl
      \off{3827636}
      \,D_{d_{4}d_{5}}^{
           d_{17}d_{11}}\,D_{d_{16}d_{17}}^{
                             d_{18}d_{12}}\,
      D_{d_{18}d_{6}}^{
         d_{14}d_{13}}\,C_{d_{19}}\,E_{d_{7}d_{1}}^{
      d_{15}d_{19}d_{8}}
      -42\,f_{d_{14}d_{15}}^{
              d_{9}}\,\nl
      \off{3827636}
      D_{d_{2}d_{3}}^{
         d_{16}d_{10}}\,D_{d_{4}d_{5}}^{
                           d_{18}d_{11}}\,D_{d_{16}d_{6}}^{
      d_{17}d_{12}}\,D_{d_{18}d_{17}}^{
                        d_{14}d_{13}}\,C_{d_{19}}\,
      E_{d_{7}d_{1}}^{
         d_{15}d_{19}d_{8}}
      -168\,\nl
      \off{3827636}
      f_{d_{14}d_{15}}^{
         d_{9}}\,D_{d_{2}d_{3}}^{
                    d_{16}d_{10}}\,D_{d_{16}d_{4}}^{
      d_{17}d_{11}}\,D_{d_{17}d_{5}}^{
                        d_{18}d_{12}}\,D_{d_{18}d_{6}}^{
      d_{14}d_{13}}\,C_{d_{19}}\,E_{d_{7}d_{1}}^{
      d_{15}d_{19}d_{8}}\nl
      \off{3827636}
      +84\,D_{d_{1}d_{2}}^{
              d_{14}d_{11}}\,D_{d_{3}d_{4}}^{
                                d_{17}d_{12}}\,
      D_{d_{5}d_{6}}^{
         d_{18}d_{13}}\,D_{d_{14}d_{15}}^{
                           d_{16}d_{10}}\,D_{d_{17}d_{18}}^{
      d_{19}d_{9}}\,D_{d_{19}d_{7}}^{
                       d_{15}d_{8}}\,C_{d_{16}}\nl
      \off{3827636}
      +56\,D_{d_{1}d_{2}}^{
              d_{14}d_{11}}\,D_{d_{3}d_{4}}^{
                                d_{17}d_{12}}\,
      D_{d_{5}d_{6}}^{
         d_{19}d_{13}}\,D_{d_{14}d_{15}}^{
                           d_{16}d_{10}}\,D_{d_{17}d_{18}}^{
      d_{15}d_{9}}\,D_{d_{19}d_{7}}^{
                       d_{18}d_{8}}\,C_{d_{16}}\nl
      \off{3827636}
      +168\,D_{d_{1}d_{2}}^{
               d_{17}d_{11}}\,D_{d_{3}d_{4}}^{
                                 d_{18}d_{12}}\,
      D_{d_{5}d_{6}}^{
         d_{19}d_{13}}\,D_{d_{14}d_{15}}^{
                           d_{16}d_{10}}\,D_{d_{17}d_{18}}^{
      d_{15}d_{9}}\,D_{d_{19}d_{7}}^{
                       d_{14}d_{8}}\,C_{d_{16}}\nl
      \off{3827636}
      +224\,D_{d_{2}d_{3}}^{
               d_{14}d_{10}}\,D_{d_{4}d_{5}}^{
                                 d_{17}d_{11}}\,
      D_{d_{14}d_{15}}^{
         d_{16}d_{9}}\,D_{d_{17}d_{1}}^{
                          d_{18}d_{8}}\,D_{d_{18}d_{7}}^{
      d_{19}d_{13}}\,D_{d_{19}d_{6}}^{
                        d_{15}d_{12}}\,C_{d_{16}}\nl
      \off{3827636}
      +84\,D_{d_{2}d_{3}}^{
              d_{16}d_{11}}\,D_{d_{4}d_{5}}^{
                                d_{17}d_{12}}\,
      D_{d_{14}d_{1}}^{
         d_{15}d_{10}}\,D_{d_{16}d_{17}}^{
                           d_{18}d_{9}}\,D_{d_{18}d_{7}}^{
      d_{19}d_{8}}\,D_{d_{19}d_{6}}^{
                       d_{14}d_{13}}\nl
      \off{3827636}
      \,C_{d_{15}}
      -126\,D_{d_{2}d_{3}}^{
               d_{16}d_{11}}\,D_{d_{4}d_{5}}^{
                                 d_{18}d_{12}}\,
      D_{d_{6}d_{7}}^{
         d_{19}d_{13}}\,D_{d_{14}d_{1}}^{
                           d_{15}d_{10}}\,D_{d_{16}d_{17}}^{
      d_{14}d_{8}}\nl
      \off{3827636}
      \,D_{d_{18}d_{19}}^{
           d_{17}d_{9}}\,C_{d_{15}}
      -140\,D_{d_{2}d_{3}}^{
               d_{16}d_{11}}\,D_{d_{4}d_{5}}^{
                                 d_{18}d_{12}}\,
      D_{d_{14}d_{1}}^{
         d_{15}d_{10}}\,D_{d_{16}d_{6}}^{
                           d_{17}d_{13}}\nl
      \off{3827636}
      \,D_{d_{17}d_{19}}^{
           d_{14}d_{9}}\,D_{d_{18}d_{7}}^{
                            d_{19}d_{8}}\,C_{d_{15}}
      +336\,D_{d_{2}d_{3}}^{
               d_{16}d_{11}}\,D_{d_{4}d_{5}}^{
                                 d_{18}d_{12}}\,
      D_{d_{14}d_{1}}^{
         d_{15}d_{10}}\nl
      \off{3827636}
      \,D_{d_{16}d_{17}}^{
           d_{14}d_{9}}\,D_{d_{18}d_{7}}^{
                            d_{19}d_{8}}\,D_{d_{19}d_{6}}^{
      d_{17}d_{13}}\,C_{d_{15}}
      +56\,D_{d_{2}d_{3}}^{
              d_{16}d_{11}}\,D_{d_{4}d_{5}}^{
                                d_{19}d_{12}}\,
      \nl
      \off{3827636}
      D_{d_{14}d_{1}}^{
         d_{15}d_{10}}\,D_{d_{16}d_{17}}^{
                           d_{18}d_{9}}\,D_{d_{18}d_{6}}^{
      d_{14}d_{13}}\,D_{d_{19}d_{7}}^{
                        d_{17}d_{8}}\,C_{d_{15}}
      -56\,D_{d_{2}d_{3}}^{
              d_{17}d_{10}}\,D_{d_{4}d_{5}}^{
                                d_{18}d_{11}}\nl
      \off{3827636}
      \,D_{d_{14}d_{15}}^{
           d_{16}d_{9}}\,D_{d_{17}d_{7}}^{
                            d_{15}d_{13}}\,
      D_{d_{18}d_{1}}^{
         d_{19}d_{8}}\,D_{d_{19}d_{6}}^{
                          d_{14}d_{12}}\,C_{d_{16}}
      -224\,D_{d_{3}d_{4}}^{
               d_{16}d_{10}}\,\nl
      \off{3827636}
      D_{d_{14}d_{2}}^{
         d_{15}d_{9}}\,D_{d_{16}d_{1}}^{
                          d_{17}d_{8}}\,D_{d_{17}d_{7}}^{
      d_{19}d_{13}}\,D_{d_{18}d_{5}}^{
                        d_{14}d_{11}}\,D_{d_{19}d_{6}}^{
      d_{18}d_{12}}\,C_{d_{15}}
    \)
  \)
  /211680
\]
\Nl}$$
\vfil\supereject
\bye